\documentclass{aastex631}
\usepackage{upgreek}

\shorttitle{The Case for Technosignatures}
\newcommand{\PSUAA}{Department of Astronomy \& Astrophysics, 525 Davey Laboratory, The Pennsylvania State University, University Park, PA, 16802, USA}
\newcommand{\PSUCEHW}{Center for Exoplanets and Habitable Worlds, 525 Davey Laboratory, The Pennsylvania State University, University Park, PA, 16802, USA}
\newcommand{\PSETI}{Penn State Extraterrestrial Intelligence Center, 525 Davey Laboratory, The Pennsylvania State University, University Park, PA, 16802, USA}

\newcommand{\SETI}{SETI Institute, 339 Bernardo Ave., Suite \#200, Mountain View, CA 94043}

\newcommand{\Ntech}{\ensuremath{N(\mathrm{tech})}}
\newcommand{\Nbio}{\ensuremath{N(\mathrm{bio})}}
\begin{document}

\title{The Case for Technosignatures: Why They May Be Abundant, Long-Lived, Highly-Detectable, and Unambiguous}

\correspondingauthor{Jason Wright}
\email{astrowright@gmail.com}

\author[0000-0001-6160-5888]{Jason T.\ Wright}
\affil{\PSUAA}
\affil{\PSUCEHW}
\affil{\PSETI}

\author[0000-0003-4346-2611]{Jacob Haqq-Misra}
\affiliation{Blue Marble Space Institute of Science, Seattle, WA, USA}
\author[0000-0002-4948-7820]{Adam Frank}
\affiliation{Department of Physics and Astronomy, University of Rochester, Rochester NY 14620, USA}
\author[0000-0002-5893-2471]{Ravi Kopparapu}
\affiliation{NASA Goddard Space Flight Center, 8800 Greenbelt Road, Greenbelt, MD 20771, USA}
\author[0000-0002-2685-9417]{Manasvi Lingam}
\affiliation{Department of Aerospace, Physics and Space Sciences, Florida Institute of Technology, Melbourne FL 32901, USA}

\author[0000-0001-7057-4999]{Sofia Z. Sheikh}
\affiliation{\SETI}

\begin{abstract}

The intuition suggested by the Drake Equation implies that technology should be less prevalent than biology in the galaxy. However, it has  been appreciated for decades in the SETI community that technosignatures could be more abundant, longer-lived, more detectable, and less ambiguous than biosignatures.

We collect the arguments for and against technosignatures' ubiquity, and discuss the implications of some properties of technological life that fundamentally differ from non-technological life, in the context of modern astrobiology: it can spread among the stars to many sites, it can be more easily detected at large distances, and it can produce signs that are unambiguously technological. 

As an illustration in terms of the Drake Equation, we consider two Drake-like equations, for technosignatures (calculating \Ntech) and biosignatures (calculating \Nbio). We argue that Earth and humanity may be poor guides to the longevity term $L$, and that its maximum value could be very large, in that technology can outlive its creators and even its host star. We conclude that while the Drake equation implies that $\Nbio \gg \Ntech$, it is also plausible that $\Ntech \gg \Nbio$. 

As a consequence, as we seek possible indicators of extraterrestrial life, for instance via characterization of the atmospheres of habitable exoplanets, we should search for both biosignatures and technosignatures. This exercise also illustrates ways in which biosignature and technosignature searches can complement and supplement each other, and how methods of technosignature search, including old ideas from SETI, can inform the search for biosignatures and life generally.

\end{abstract}

\keywords{}

\section{Introduction} \label{sec:intro}

\subsection{The Development of Biosignatures and Technosignatures}

The search for life elsewhere in the universe focuses on identifying biosignatures that would indicate the presence of extraterrestrial life. An extension of the search for biosignatures is the search for ``technosignatures'' \citep{tarter06}, which specifically focuses on identifying remotely observable signatures of extraterrestrial \textit{technology}.

The search for remotely detectable biosignatures has developed greatly since the discovery of exoplanets \citep{Wolszczan92,Mayor_queloz}. Since that event, the field has tackled a number of scientific questions and challenges including the coupling of climate models to atmospheric chemistry networks for the production of synthetic observations \citep{seager2012astrophysical, Kaltenegger2017, Schwieterman2018, Meadows2018b, Catling2018, Walker2018, Fujii2018, lammer2019role, Grenfell2017}, the exploration of alternative pathways for biological processes such as photosynthesis \citep{WR02,KSA07,Schwieterman2018,LL20} and the ongoing development of agnostic biosignatures such as markers of chemical complexity \citep{JAG18,Walker2018,MMC21}.  

Searches for remotely detectable technology have benefited from renewed energy in recent years due to a confluence of several events, including: the discovery that habitable zone rocky planets are ubiquitous in the universe \citep{Burke2015, DC2015, mulders2018, Hsu2018, Kopparapu2018, bryson2021}, the resulting development of a robust astrobiology program focused on biosignatures, the Breakthrough Listen Initiative---which has pledged \$100 million over 10 years towards the search for technosignatures \citep{Worden2017}---and a renewed interest from the US Congress and NASA in funding technosignature research. This renewed investment has cultivated a much broader range of avenues for the detection of extraterrestrial life. 


Both fields (technosignature and biosignature searches) have developed expectations and theoretical frameworks for the abundance of signs of life in the universe. Our goal here is to perform a \textit{comparison} of these expectations. Of course, a objective, quantitative comparison of the actual relative abundances of technosignatures and biosignatures is difficult because it depends on details of extraterrestrial life that we cannot know for certain until we have some examples to learn from. 

To make our comparison, we will draw on many arguments that are familiar to veterans of SETI programs, but have not, in our experience, had much purchase in the broader astrobiology landscape. This is likely because many of these arguments have appeared in journals rarely read by astrobiologists, in conference proceedings, in books, and in the gray literature---that is, where they have been formally published at all.  Our goal is to compile these arguments into a single coherent argument for the plausibility of the abundance of technosignatures, and to do so in the context of modern astrobiology (especially since many of these arguments were first made before that field even existed in its modern form).


The essence of our argument is that technology---and its attendant technosignatures---differs in \textit{fundamental} ways from biology---and its biosignatures---in that it can spread far beyond its origin in space, time, and scope \citep[e.g.,][]{WHK80,MDP82,RAF83}. In this work, we explore how this difference gives technosignatures almost unlimited potential for key metrics associated with exo-life searches: \textbf{abundance; longevity; detectability;} and \textbf{ambiguity}.\footnote{The Nine Axes of Merit for Technosignature Searches \citep[][]{Sheikh2020} provide a framework for a similar comparison \textit{within} the technosignature field. ``Abundance'' here maps loosely to the ``Inevitability'' axis in that work, and ``Longevity'' here to the ``Duration'' axis, while Detectability and Ambiguity are more directly comparable.} 

This is important because the most commonly used heuristics for describing biosignatures---such as the Drake Equation---can lead one to the erroneous conclusion that technosignatures \textit{must} be poorer search targets than biosignatures. As we will see, there is no incontrovertible reason that technology could not be more abundant, longer-lived, more detectable, and less ambiguous than biosignatures.

\subsection{The Role of the Drake Equation}

For the first two of our four points, abundance and longevity, we can turn to the heuristic of the Drake Equation \citep{DrakeEquation,VDD15}, which has for decades been a powerful guide to how to think about the prevalence of life, biosignatures, and technosignatures. Usually expressed in its canonical form:
\begin{equation}
    N = R_* f_p n_p f_l f_i f_c L 
\end{equation}
\noindent it seeks to express an estimate of the number of planets with communicative species in the Milky way. 

Broadly speaking, the first three terms on the right-hand side of the equation, capturing the rate of star formation in the Galaxy ($R_*$), the fraction of stars with planets ($f_p$), and the number of habitable planets per star ($n_p$), can be determined by astronomers without having to rely on much interdisciplinary input, and have reasonably well-known values. The next term, $f_l$, the fraction of those habitable planets that develop life, helps determine how likely it is that searches for biosignatures succeed. The next two, $f_if_c$ capturing the fraction of \textit{those} planets that give rise to intelligence and communicative species, respectively, and so narrow things down further to the fraction we could detect via communicative signals, such as radio waves. $L$ is the average length of time a species spends sending a detectable communicative signal, yielding an expectation value for the number of communicative species at any moment.

If we think of technosignatures more broadly and do not just focus on communication, we can write a pair of Drake-like equations for biosignatures and technosignatures generally. The first

\begin{equation}
    \Nbio = R_* f_p n_p f_l L_b
\end{equation}

\noindent captures the number of planets with detectable biosignatures, where $f_l$ here is interpreted as the fraction of planets that develop sizable biospheres of the sort that give rise to remotely detectable biosignatures (and, potentially, technological life) and $L_b$ is the average length of time that biosphere remains detectable.

Similarly, we write
\begin{equation}
    \Ntech = R_* f_p n_p f_l f_t L_t 
\end{equation}
\noindent which captures the number of planets with detectable technosignatures, where here $f_t$ is the fraction of planets with life with detectable technosignatures of any sort and corresponds to $f_if_c$ in the equation's canonical form, and $L_t$ is the length of time that such technosignatures remain detectable. 

\subsection[]{The case for $\Nbio\gg\Ntech$, and the flaw in that argument}


A na\"ive conclusion from the Drake Equation is that $\Nbio \gg \Ntech$, based on at least two seemingly sound assumptions. The first assumption is that $f_t \ll 1$, since the development of technology is presumably a rare event \citep[e.g.][]{GGS64,Mayr85,EM04}. The second assumption is that any technological life that \textit{does} arise will exist for only a small fraction of the time that life of any kind exists on the planet \citep[e.g.,][]{SVH61,PAS80}. Both of these assumptions are based on the sole example of the evolutionary history of Earth. 

In other words, these two assumptions would lead to the expectation that
\begin{equation}
    \frac{\Ntech}{\Nbio} = f_t\frac{L_t}{L_b} \ll 1
\end{equation}

The primary flaw in the reasoning from the Drake Equation is that it restricts technology to a small subset of biospheres in time and space, and that it does so \textit{tacitly} and \textit{by assumption}. In truth, while technology may \textit{originate} from biology, it may also exist \textit{independently} of it in many ways that the Drake Equation does not capture.

To give some examples of what we mean:
\begin{enumerate}
    \item Technology can long outlive the biology that created it \citep[e.g.,][]{DLH91,Cirkovic19}. One can thus have technosignatures without technological life, and even without a biosphere.
    \item Technological life can spread to other worlds, and thus create sites of technosignatures that outnumber biospheres \citep[e.g.,][]{WHK80,MDP82,MDP83,RAF83}.
    \item Once created, technology could self-replicate and spread without any further association with life at all \citep[e.g.,][]{tipler80,AS13,JG16}.
    \item Technosignatures can exist beyond planets, for instance arising from spacecraft \citep[e.g.,][]{VHP77,MJH86,GC13,ML20}.
\end{enumerate}


\section{Abundance}


Consider the Solar System, which at the moment hosts only a single planet with remotely-observable biosignatures (at least, any obviously discernible ones) but has \textit{four} major planets with potentially detectable technosignatures. In addition to Earth, Jupiter and Venus are both orbited by probes actively transmitting radio waves, and Mars hosts several orbiters, landers, and rovers.

Thus, in our Solar system,  we have \Nbio=1, \Ntech=4. The radio technosignatures of these other planets are obviously weak at the moment, but the strength of such technosignatures will increase if and when space exploration of these planets intensifies in the future. There are substantial plans to increase humanity's presence on Mars, and regardless of whether it takes decades or millennia to realize those visions, it may be reasonable to expect Mars at some point to have strong, remotely detectable radio and other technosignatures, while still lacking a conspicuous indigenous biosphere and its attendant biosignatures.

Looking further, there is no reason that future humans or an alien species could not choose to use some uninhabitable worlds as sites for significant technology, whether for industry, energy generation or even waste processing, creating entire ``technospheres'' spatially independent of the biosphere which spawned them.  These ``service worlds,'' described by Lingam et al. \textit{in prep.}, may might not host any biology at all, and might outnumber sites of biology from their originating species.

There is also no reason to think that technological life in the Galaxy cannot spread beyond its home planetary system \citep[cf.][]{MB65,FDD80}. While interstellar spaceflight of the sort needed to settle a nearby star system is beyond humanity's current capabilities, the problem is one being seriously considered now, and there are no real physical or engineering obstacles to such a thing happening (e.g., \citealt{JHM92,Ashworth2012,ML21}). Even if we cannot envision it happening for humans in the near future, it is not hard to imagine it transpiring in, say, 10,000 or 100,000 years. Many \citep[e.g.][]{Purcell1960} have argued that such travel is unlikely because, while not \textit{impossible} it certainly seems difficult, and so radio transmissions are a more likely and energy-efficient form of contact. But the two are not mutually exclusive activities, and energetics are not as daunting as might seem at first glance: \citet{Hansen2021} show that there are episodes of close encounters between stars that can lower launch costs by an order of magnitude below their mean values, and \citet{Carroll19} show that such close encounters can drive an efficient settlement of the Galaxy even for slow ships with short range.

Therefore, the idea that such spreading would happen is not outrageous---indeed, it lies at the very heart of the Fermi Paradox, which \textit{assumes} that technological life will spread throughout the Galaxy \citep{hart75,tipler80,FermiParadox}.\footnote{Indeed, Hart and Tipler took the opposite extreme, arguing that the spreading would be so extensive that it would be too obvious to ignore, and so therefore has never happened. See references in \citet{Carroll19} for many examples of rebuttals to this reasoning.} Because the time it would take even slow ships to cross the Galaxy is small compared to its age, a settling species has had ample time to settle every star in the Galaxy. \citet{Carroll19} and others cited therein have shown that there are a wide range of rather conservative settlement parameters (e.g.\ very infrequent launches of ships only capable of reaching the nearest stars) that lead to exponential growth of the number of settlements, limited only by the number of suitable stars.

If such spreading is common, then the Drake Equation must \textit{underestimate} \Ntech\ by a very large factor, potentially even up to order $10^{10}$ in some extreme scenarios. Indeed, the idea of including a spreading factor in the Drake equation has been suggested by many authors \citep[see, e.g.][]{WHK80,Prantzos2013}. 

Does the same logic apply to \Nbio? If the question is posed purely in terms of nontechnological life, then the answer would appear to be ``no'', unless interstellar transport of life is highly effective \citep[e.g.][]{crick1973directed,napier2004mechanism,grimaldi2021feasibility,LGB22}. In addition, however, \Nbio\ should include biosignatures stemming from biospheres that have been purposely engineered for habitation, for example, planets which have been ``terraformed''.

Such spreading of biosignatures is plausibly a much rarer event than the spreading of technosignatures; the problem of how to have robots and self-replicating machines settle a nearby star system would likely be easier to solve than the problem of how to send a crewed ship, because one need not worry about safety and life support for the crew during the trip \citep{Freitas80probes,tipler80,BH21}. This case may, however, be overstated in actuality. The spread of life need not invoke enormous ``generation ships'' \citep{HPB12} or warp drives: one can imagine many solutions to the difficulties of the very long journey between stars. For intsance, one might imagine cryostasis or even the re-generation of life ``from scratch'' by using artificial ova and DNA sequences stored in computer memory \citep[e.g.,][]{RAF83,CHH12,HLL18,MRE21}, and \citet{Hansen2021} discuss how stellar motions mean that occasionally such trips can be quite short. Regardless, once it arrives at a new home, such life will almost certainly create technosignatures (since it used technology to get there), and some fraction of them may also eventually give rise to a new biosphere, as well.
 
Finally, and perhaps most important for the issue of abundance, one can have many more sites of technology than technospheres, because one technosphere can produce myriad technosignatures that spread across space.  A simple example already raised is the launching of interstellar vessels for any purpose, whether for settlement, scientific studies, simply trade, or something else. On a related note, self-replicating probes are already not far removed from the technological capabilities of humans \citep{BH21}. One can, therefore, imagine a single example of technological life producing a single example of a technosphere which would, nonetheless, produce an arbitrary number of technosignatures distributed across interstellar space. In terms of the Drake Equation, we might write that as $f_t \gg 1$.

Thus while the Drake Equation may underestimate \Nbio\ to some degree, there are reasons inherent to technology that permit us to argue that the attendant degree of underestimation will be much larger for \Ntech, and by potentially enormous factors.

\section{Longevity}\label{SecLongevity}

The next place the Drake Equation implies low values of \Ntech/\Nbio\ is in $L_t$, which one might think must be much smaller than $L_b$. In particular, $L_b$ for Earth is of order 4 Gyr, and $L_t$ seems to be of order decades to millennia, based on human technosignatures. Much ink has also been spilled tying the $L_t$ in the Drake Equation to fears of human catastrophes, for instance from nuclear or biological warfare or climate change, and arguing it would be optimistic to think it could be orders of magnitude larger than a thousand years or so; indeed, such arguments are as old as the Drake Equation itself \citep[e.g.,][]{DrakeEquation,MB65,shklovskii1966,Gott93,charbonneau2022mixed}.

But there are at least five strong reasons to think $L_t$ might be greater than $L_b$.

\subsection[]{Humanity may be a poor guide to $L_t$}

Even if we choose to focus just on technological life and the technospheres they create we must be careful of anthropocentrism, which is one of the trickiest traps to escape in the theory of technosignature searches. Humanity is the only example we have of the sort of species we seek to detect via technosignatures, and so it is both an essential and misleading data point. 

One critique of this approach to technosignature searches is that it seeks technosignatures of the sort that humanity happens to create now, ignoring that other societies might develop radically different technologies and behaviors than we have (see \citealt{KD11}), especially if they have been doing so for far longer than humans have. And indeed, we should avoid projecting human values and tendencies (or what we \textit{imagine} human values and tendencies to be) onto alien species (e.g., \citealt{CTR01}). But this criticism is equally just as true when assigning \textit{small} values to $L_t$ because one is pessimistic about humanity's future. In actuality, there is likely some wide distribution of values for $L_t$ for alien species, including very large values for some of them \citep{KFS20,BalbiCirkovic2021}.

In short, there is no reason to think that humanity's value for $L_t$ will be characteristic of alien societies, which anyway presumably occupy a wide range, and whose mean may be driven by a long tail out to large values \citep[e.g.,][]{shklovskii1966}.

\subsection[]{We cannot use the past to predict $L_t$ for Earth's future}

Even if we do choose to base our expectations of $L_t$ on humanity---since it is, after all, the only example we have---we should still be skeptical that $L_t\ll L_b$ because this intuition is based only on Earth's \textit{past}, which may be a poor guide to the future where technology is involved. 

Humanity's impact on many of Earth's geological and biological processes has been profound, and some have suggested that the emergence of technology actually heralds a new state of Earth that will persist as long as the Earth does \citep{baum2019long}; see, for instance, the Sapiezoic Era \citep{Grinspoon2019} or Earth as a Class {\sc v} planet \citep{Frank18}. Of course, such profound changes \textit{could} be self-limiting, as humanity may face severe global catastrophic risks \citep{BC08} or existential risks \citep{bostrom2002existential} that could prevent a long-term sustainable future. Examples of such self-limiting threats include climate change, nuclear war and pandemics.

But, as pointed out by \citet{DrakeEquation}, unless such a catastrophe is apocalyptic, it will not necessarily end our technosignatures. The collapse of civilizations and even widespread ecological devastation would certainly be deeply terrible, but unless they result in \textit{human extinction}, humanity may be able to eventually recover, even though the ecological consequences might extend to millions of years \citep{MK01}. Catastrophic events conceivably lead to scenarios in which the loss of infrastructure, knowledge, and resources is so great that recovery of large-scale technology is impossible \citep{baum2013double}. But even then, what humans have developed once, they could potentially develop again, even from scratch.\footnote{However, the subsequent development of widespread technology could be impaired by prior technological life depleting a planet's easily-accessible resources, if the time-scales of development and resource replenishment are comparable.} 

And even in the face of catastrophe, humanity may retain much of its knowledge and enough infrastructure to quickly be able to recover any lost technology, for example, by establishing refugia. In other words, the timescales for catastrophe and the resulting recovery might be short compared to the future of Earth's habitability, so while such catastrophes may lower the duty cycle during which our technosignatures are detectable, they do not necessarily present a hard upper limit to $L_t$. 

Likewise, humanity is the first species on Earth that can \textit{prevent} its own extinction with technology, for instance by diverting asteroids, stopping or mitigating pandemics, or building ``lifeboat'' settlements elsewhere in the Solar System or beyond \citep{baum2015isolated,TG17,TD18}. This means that the upper limit on our technology's survival is essentially unlimited in theory, even in the face of inevitable natural catastrophes. Apart from these modern examples, Earth-analogs from human history teach us that a technological downshift---to temporarily become less technological until circumstances improve---is a common and healthy adaptation to catastrophe in human history, and that technology and longevity are in this way inextricably linked \citep[][]{denning2011earth}.

Further, $L_b$ in the past is of order a few Gyr, but Earth's future probably has $\lesssim 1$ Gyr of habitability  \citep{caldeira1992life,WT15,OR21}, putting a hard limit on $L_b$.  Earth's technosignatures, however, have no such limits except those set by geological processes on functionality, since even an uninhabitable Earth can still host technology (as Mars does). Indeed, considering the Solar System as a whole, there is no reason technology could not exist for several Gyr, meaning that it is plausible that $L_t > L_b$ might be achievable for the Solar System.  


In short, our expectations are that $L_t \ll L_b$ should not be based on Earth's past, but on its future, and that the probability distribution of $L_t/L_b$ in this future \textit{even for human technology} has a long tail extending to large values.

\subsection[]{Other species and $L_t$ for Earth life}

Even if we accept that Earth life is a good guide to extraterrestrial life \textit{and} that humanity's future is short, we can \textit{still} argue that $L_t$ could be quite large because Earth has a plausible chance of giving rise to \textit{another} technological species after humanity's extinction \citep[pg. 328]{DrakeEquation}; see, however, \citet{TH22}. After all, few of the most striking traits of our species are altogether unique in the animal world: intelligence, tool-use, communication, and social behavior are all common \citep[as reviewed in][Chapter 3]{ML21}. Given that it has already happened (at least) once, there is no inexorable reason to think that as long as animal life exists at all, another species cannot develop an even longer-lived technological society. The actual probability of such a second evolution of technology would, of course, strongly depend on the extent of the particular extinction event and whether the evolution of technological intelligence is ``favored'' by evolution (see \citealt{RP20}).

Pessimism about humanity's future is thus not sufficient to argue that $L_t$ for Earth is short, especially if one believes that our propensity for self-destruction or environmental degradation is a particularly human failure.

\subsection[]{We do not know $L_t$ for Earth's past}

In a similar vein, if we grant that other terrestrial species could be capable of technology, then we cannot even use Earth's past to precisely constrain $L_t$ empirically because we do not know it! Our assumption that $L_t\ll L_b$ is usually based on what we \textit{think} we know about Earth, namely, that remotely-detectable technosignatures on Earth have only existed for decades or, optimistically, perhaps in the millennia since humans developed agriculture---that is, $L_t/L_b \sim 10^{-7}-10^{-6}$. 

This assumption seems well justified, but a little examination reveals that we actually have very little to base it on. \citet{Schmidt19} showed that, while humanity's impact on the geological record (marking the Anthropocene epoch) will be indelible, the fact that it is due to technology will be indiscernible on a timescale of millions of years. That is to say, the Anthropocene will be obvious over geological timescales as an important event in Earth's past, but there will be little to no evidence that it was the result of technology, as opposed to being a purely natural phenomenon. The conclusion is that if there have been prior episodes of technological life on Earth, \textit{we would not know it}. 

So while Earth's past certainly teaches that $L_t< L_b$, the magnitude of the difference could easily be much smaller than we might guess.  If there have been tens of millions of years of prior technology, and if this technology could be conceivably detectable, then $L_t/L_b \sim 10^{-2}$. We just don't know.

To summarize Section \ref{SecLongevity} so far, using Earth as a guide for our expectations for $L_t/L_b$ is probably unreliable because we do not know $L_t$ for Earth's past; and even if we did we should not use it to predict $L_t$ for Earth's future; and even if we did, we should not expect Earth to be a good guide for alien life; and even if we did, we should expect a broad distribution of longevities across alien species. 


\subsection[]{Technologies and technosignatures can outlast technological life}

An additional feature of technology is that it can be robust enough to outlive its creators, and even the conditions for biology on the planets that created it. Another important difference between technology and biology is that while we would not expect biosignatures to survive the end of their biosphere, we cannot say the same for technosignatures. 

To give just one example, if a fraction of a planet were covered with solar collectors \citep{Lingam17}, then the associated technosignature in the form of a reflectance spectral imprint could survive long after after its creators or even the entire biosphere were to end.  If the solar collectors were placed on an uninhabited world such as airless moon or planet, the timescale associated with the persistence of these technosignature may perhaps run into the hundreds of millions of years. In addition, automation and artificial intelligence imply that technologies associated with planetary technospheres or surviving in space can persist in a functioning state long after the technological life which created them ceases to exist. It is plausible, for instance, that technology in space (even if defunct) might survive over long timescales, and searches for such technosignatures have been proposed \citep{DLH91,Carrigan10}.

\section{Detectability}
\label{sec:strength}


There is a maximum strength of any exoplanetary biosignature: life on another planet cannot modify a space larger than its own biosphere---that is, barring unproven mechanisms that could move microbial life among planets. Life also generally does not appear to concentrate and radiate power or to tightly concentrate it temporally or spectrally \citep[cf.][]{DMR92}, implying that it can only be detected passively on exoplanets, for instance via atmospheric gases or surface reflection spectra.

Technosignatures, by contrast, have essentially no upper limit to their detectability. \citet{kardashev64} outlined a famous scale spanning over 20 orders of magnitude of potential amounts of power available for technological manipulation, from humanity's current energy supply today to the total stellar luminosity of a galaxy. 

Indeed, humanity's technosignatures are \textit{already} on par in strength with Earth's biosignatures, and may even exceed them in detectability at interstellar distances. Humanity does not currently have the technology to detect most of Earth's biosignatures if seen from the distance of, say, $\upalpha$ Centauri. It is even unclear if JWST will provide that capability, although it may be able to detect potential biosignatures on the TRAPPIST-1 planets \citep{Adam2018, Thomas2019, Jake2019}. 

On the other hand, when completed, the full SKA may be sensitive enough to detect the combined chatter of our aircraft and other radar at a few parsecs \citep[cf.][]{Loeb2007,Forgan2011}. \citet{Kopparapu2021} showed that LUVOIR might be able to detect NO$_\mathrm{x}$ in an Earth-like planet at 10 pc, in principle. Furthermore, Haqq-Misra et al. (2021, submitted) estimated that some chloroflorocarbons (CFCs) might be detectable, if present, on TRAPPIST-1\textit{e} with JWST, in principle.

So, while it is unclear as to which would be more detectable on a putative technological species' home planet, the lesson from Earth is that biosignatures and technosignatures may be of quite similar strengths.\footnote{Intentional technosignatures (like the Arecibo message), which we have not outlined, are detectable by roughly Earth-level technology over much larger distances of order 1 kpc \citep{Tarter01}, but are extremely intermittent and narrow band, so perhaps less detectable than our biosignatures in that specific sense.} It is also not unreasonable to argue that, for an older society with more time for its technology to have grown, technosignatures might be much stronger, and so have the upper hand. And on the previously adumbrated ``service worlds'' and other places technology has spread, we would expect technosignatures to be entirely dominant.

On balance, the lesson from Earth is that we expect technosignatures to be at least as detectable---and potentially greatly more so---than biosignatures on any planet with a technological species.

\section{Ambiguity}

Ambiguity is a major problem with searches for both biosignatures and technosignatures. This comes in two forms: ambiguity in the \textit{nature} of a detection (e.g., is this spectral signature really methane / phosphine / NO$_\mathrm{x}$?) and ambiguity in the meaning of the detection (e.g., could those species be abiogenic?).

Technosignatures, however, offer at least the potential for complete unambiguity \citep[e.g.][]{Margot2019}. A narrowband radio signal (of order $\sim$ Hz spectral width) can \textit{only} be technological \citep[e.g.][]{Tarter01}. Indeed, the only ambiguity comes from identifying the \textit{source} of the technology, since it could be human, as is often the case with detections in radio SETI \citep[see the recent example of ][]{sheikh2021analysis}. The case for an artificial gas might be slightly more ambiguous because an exotic biology might create species we think of as purely artificial, but the primary confounders then are biogenic sources, which is still a ``win'' for astrobiology.

Other technosignatures, such as the waste heat of industry, have many astrophysical confounders and are thence strongly ambiguous, but such is also the case for many biosignatures. One possible solution in these cases is to look for other potential technosignatures (comprising ``constellations'') that could increase our confidence and build a contextual understanding of the detection, rather than relying on a single observed feature at a given wavelength. Similar strategies are adopted to remotely detect biosignatures on exoplanets, where multiple biologically produced gases (gas `fluxes') are used to obtain an understanding of the planetary environment under which such detections would be made \citep{Catling2018,Schwieterman2018}.

Finally, for both biosignatures and technosignatures, we may fail to recognize what we are seeing because it is so unexpected, and therefore mistake it for a strange, natural and purely abiogenic phenomenon. So, on balance, it is unclear whether biosignatures or technosignatures would be more ambiguous, because it will depend strongly on \textit{which} signature or collection of signatures is being considered, but there exist at least some technosignatures for which ambiguity is simply not a problem.

\section{Discussion and Conclusions}

We have evaluated four ways to assess the relative prevalence and detectability characteristics of biosignatures and technosignatures in the search for life in the universe, and find that in all cases it is unclear which of the duo would be stronger, and that technosignatures have a higher ceiling.

The primary reasons in the first two cases are related to the assumptions baked into the Drake Equation, namely, that life stays put and lasts only as long as its biosphere. To put it differently we can state that life and technology have related but potentially independent evolutionary trajectories.  Alternatively, to put it another way, biology, which gives rise to biosignatures in a biosphere, has different spatial and temporal limitations than technology, which gives rise to technosignatures both within and beyond a technosphere.

This means that while the \textit{emergence} of biosignatures and technosignatures on a planet can be seen as complementary aspects of the same problem, their evolutionary trajectories need not be co-extensive for all time. Processes which produce biosignatures and those that produce technosignatures may begin sharing the same planet with a certain distribution and strength. But over time their distribution and strengths may diverge by orders of magnitude. As the field of technosignature search matures, we develop new considerations of the forms such divergences can take \citep[e.g., ][]{Bradbury11,WrightExoplanetsSETI,Lingam2019,BalbiCirkovic2021,JTW21}. These new theories offer insights into the possibilities for life in the Universe, as well as an opportunity to reassess priors in terms of search strategies for biosignatures and technosignatures.

Because the Drake Equation implicitly assumes technology follows a biology-like evolutionary trajectory, it may not accurately represent the prevalence of detectable technosignatures.  In other words, it ignores the very real possibility that technological life can spread efficiently \citep{WHK80}. In particular, while a na\"ive application of this formalism might argue for there being many more sites of biosignatures than technosignatures, the spread of technology could reasonably imply that the number of sites of technosignatures might be larger than that of biosignatures, potentially by a factor of as much as $> 10^{10}$ if the Galaxy were to be virtually filled with techonology.

In the case of longevity, the history of technology on Earth (or, rather, our \textit{perception} of that history) is not necessarily a good guide for the future of technology on Earth, and may not even be pertinent for the longevity of technology elsewhere in the Galaxy. In addition, there is no incontrovertible reason that technology could not far outlive its parent species, or even its parent biosphere. Combined with our preceding abundance arguments, this increases the maximum value of the number of \textit{present-day} sites of technosignatures to be many times larger than the number of sites of biosignatures, potentially by many orders of magnitude.

In the case of detection strength, we argue that the case is roughly a tie for biosignatures and technosignatures on the present-day Earth and Solar System, and that in the future, technosignatures may very well attain the more obvious presence. By this reasoning, we should expect a very long tail extending out to very strong technosignatures to be found, perhaps associated with long-lived technological species, for which the likelihood of contact is higher \citep{KFS20,BalbiCirkovic2021}. Lastly, in the case of ambiguity, we have argued that the case is again no worse than a tie, with certain technosignatures being significantly less ambiguous than any known biosignature.

It thus makes sense that the collective efforts of astrobiologists should strive to include vigorous searches for both technosignatures and biosignatures, since neither is clearly the superior search target. Another conclusion is that the two communities have much to gain from cross-fertilization of the sort we have attempted here. 

For instance, the biosignature search community is developing many mature frameworks for designing and interpreting the results of life detection experiments, including the development and selection of good biosignatures to search for, selection criteria and prioritization of target lists, and frameworks for determining the confidence of detection and dealing with ambiguity \citep[][to give a few recent examples in addition to the references in Section~\ref{sec:intro}]{Stark2019,Truitt2020,Tuchow2020,Tuchow2021,Ladder,Green2021}.  

In many cases, such as the ones we have presented here, there is substantial prior art in the SETI community, and indeed both communities are engaged in similar problems and have often come to similar conclusions from different directions. Both communities would thus benefit from comparative analyses of the kind we have performed here, and from further collaborative efforts and integration of ideas.

\begin{acknowledgements}

The Center for Exoplanets and Habitable Worlds and the Penn State Extraterrestrial Intelligence Center are supported by Penn State and its Eberly College of Science.

J.T.W., M.L., J.H.M., and A.F. gratefully acknowledge support from the NASA Exobiology program under grant 80NSSC20K0622. 

S.Z.S. acknowledges that this material is based upon work supported by the National Science Foundation MPS-Ascend Postdoctoral Research Fellowship under Grant No. 2138147

The results reported herein benefitted from collaborations and/or information exchange within NASA’s Nexus for Exoplanet System Science (NExSS) research coordination network sponsored by NASA’s Science Mission Directorate.
    
This research has made use of NASA's Astrophysics Data System Bibliographic Services.

\end{acknowledgements}


\end{document}